\documentclass[twocolumn]{jpsj3}

\usepackage{graphicx,amsmath,amssymb,bm}

\allowdisplaybreaks 

\usepackage{color}
\definecolor{Green}{rgb}{0,0.7,0}

\newcommand{\e}{ {\rm e}}
\newcommand{\ET}{ $\alpha$-(BEDT-TTF)$_2$I$_3$}

\newcommand{\bk}{ \bm{k}}
\newcommand{\bkD}{ \bm{k}_{\rm D}}
\newcommand{\eD}{ \epsilon_{\rm D}}
\newcommand{\ep}{ \epsilon }
\newcommand{\ka}{k_x}
\newcommand{\kb}{k_y}
\newcommand{\kc}{k_z}

\newcommand{\kabc}{k_x+k_y+k_z}
\newcommand{\kab}{k_x+k_y}
\newcommand{\kbc}{k_y+k_z}
\newcommand{\kac}{k_x+k_z}

\begin{document}

%-------------------------------
\title{
Anisotropic Conductivity of Nodal Line  Semimetal in Single-Component Molecular Conductor [Pd(dddt)$_2$]
}
\author{
Yoshikazu Suzumura\thanks{E-mail: suzumura@s.phys.nagoya-u.ac.jp}
}
\inst{
%$^1$
Department of Physics, Nagoya University,  Chikusa-ku, Nagoya 464-8602, Japan \\}

%\recdate{August   \;\;\;, 2017}

%%%%%%%%%%%%%%%%%%%%%
\abst{
Using a tight-binding model, we theoretically examine the 
  anisotropic conductivity of the  nodal line semimetal 
 of a  three-dimensional  Dirac electron in a single-component 
   molecular  conductor [Pd(dddt)$_2$], which  consists  of four molecules 
  with  HOMO and LUMO orbitals per unit cell. 
The conductivity shows an anisotropy given  by  $\sigma_y > \sigma_x > \sigma_z$ 
 in accordance with  that of  the velocity of the Dirac cone 
 where $z$ is the interlayer direction and $y$ is the molecular stacking  direction. 
 With increasing pressure, the nodal line semimetal emerges,  followed by   
  a loop of the Dirac point where  $\sigma_x$ takes its maximum at a pressure. 
Such a pressure dependence is studied by calculating the density of states 
   and  chemical potential.
The temperature dependence of anisotropic conductivity is examined at low temperatures 
    to obtain  
 a constant behavior, which  is ascribed to %the nodal line semimetal of 
      the Dirac electron.
The relevance of the present calculation to the experiment is discussed.
  }

%\begin{document}

\maketitle

%%%%%%%%%%%%%%%%%%%%%%%%
\section{Introduction} 
%%%%%%%%%%%%%%%%%%%%%%%%%%

 Two-dimensional massless Dirac fermions are  interesting  topics 
as shown by the quantum Hall effect in graphene. 
\cite{Novoselov2005_Nature438}
 Extensive studies on Dirac electrons have been explored  
  in molecular conductors  which are characteristic as the bulk   system.
\cite{Kajita_JPSJ2014,Kato_JACS}
In the organic conductor,\ET 
\cite{Mori1984} (BEDT-TTF=bis(ethylenedithio)tetrathiafulvalene), 
   a two-dimensional  Dirac electron with a tilted Dirac cone was found
\cite{Katayama2006_JPSJ75} 
 by noting the  vanishing  of the density of states at the Fermi surface.
\cite{Kobayashi2004} 
The Dirac point  was calculated using  the tight-binding model 
 with  the transfer energy estimated by 
 the extended H\"uckel method,\cite{Kondo2005} 
 and  the existence  was verified by 
 the electronic structure of first-principles calculation.
\cite{Kino2006} 
Such a Dirac electron explains well the transport experiment 
 under pressure.\cite{Kajita1992,Tajima2000}.

A single-component molecular conductor 
  [Pd(dddt)$_2$] (dddt = 5,6-dihydro-1,4-dithiin-2,3-dithiolate) exhibits 
Dirac electrons 
 owing to its almost constant  resistivity at low temperatures and 
under pressure.\cite{Cui} 
 The  Dirac cone located between the conduction   and 
 valence bands  was found by first-principles calculation
\cite{Tsumuraya_PSJ_2014} and  examined  
 using a tight-binding model under pressure,\cite{Kato_JACS}
 which is described by an 8 x 8 matrix Hamiltonian with the energy band 
$E_\gamma(\bk)$, [ $\gamma$ =1(top), c, 8(bottom)]. 
The crystal consists of four molecules per unit cell with the HOMO and LUMO orbitals, which are symmetric and antisymmetric around the inversion center of the Pd atom, respectively. 
In contrast to the  organic conductor,
 this material provides a line of 
 the  Dirac point in the  three-dimensional momentum space 
  $\bk =(k_x,k_y,k_z)$,  where 
 there are  transfer energies along the interlayer direction ($k_z$) in addition to  the intralayer plane of $k_x$ and $k_y$. 
At ambient pressure, the insulating state  is obtained 
  since the chemical potential, owing to the half-filled band, is located 
 in a gap   between the LUMO  and HOMO bands. 
Under high pressure, the Dirac point emerges between $E_4(\bk)$ and $E_5(\bk)$ 
when the HOMO band becomes larger than the LUMO band at 
 $\bk=0$.  
The Dirac point gives 
  a loop  in the three-dimensional Brillouin zone.
\cite{Kato2017_JPSJ}
The Dirac electron in [Pd(dddt)$_2$]  is exotic since the Dirac point 
 originates from  the combined effect of 
  the intralayer  and  interlayer transfer energies. 
 The gap function of $E_4(\bk)-E_5(\bk)$, where the zero gap gives 
 a loop of the  nodal line of the Dirac point,
   forms  a contour with a cylindrical shape   
 along the line   suggesting a nodal line semimetal 
 in three-dimensional momentum space.
\cite{Murakami2007,Burkov2011} 
However, the  property of such a  semimetal 
 is not yet clear   
    owing to   the complicated treatment of the chemical potential 
 located on the loop, 
 although isotropic property of such a system  is partly known from 
  the magnetic susceptibility.\cite{Suzumura2017_JJAP}
 Thus,  it is  of interest to examine the anisotropic behavior of 
 electric conductivity, which  originates from 
  the  three-dimensional loop.

In this study,  such a  property of the  Dirac electron is examined 
 by calculating  electric conductivity, which is useful  
 for understanding  the  experiment on resistivity under pressures.\cite{Cui}
In Sect. 2, the  model and formulation are given.
In Sect. 3, a loop of the Dirac point located 
     between the conduction and  valence bands 
       is shown for   typical pressures. 
  Anisotropic conductivity for such a semimetal is examined 
  by varying pressure. 
A maximum conductivity as a function of pressure  is obtained 
 and analyzed using 
 the density of states. The effect of temperature is also calculated.  
Section 4 is devoted to the summary and discussion.

%%%%%%%%%%%%%%%%%%%%%%%%
\section{Model and Formulation}
%%%%%%%%%%%%%%%%%%%%%%%%

 A single component molecular conductor [Pd(dddt)$_2$] 
  consists of four kinds of molecules (1, 2, 3, and 4)
   with HOMO and LUMO orbitals in the unit cell,
 \cite{Kato_JACS}
 and  has a crystal structure with two kinds of layers  where  layer 1 
   includes molecules 1 and 3, 
     and  layer 2 includes  molecules 2 and 4. 
Transfer energies, which are given by pairs between nearest-neighbor molecules,
 are as follows.  
 The interlayer  energies in the $x$-$y$ plane  are given 
    by   $a$ (1 and 2 molecules, and 3 and 4 molecules) 
             and $c$ (1 and 4 molecules, and 2 and 3 molecules). 
The intralayer  energies are given by 
   $p$ (1 and 3 molecules) and $q$ (2 and 4 molecules)
    and   $b$  (perpendicular to the $x$-$z$ plane) being the largest one. 

 On the basis of the crystal structure,\cite{Kato_JACS} 
 we examine  a tight-binding model Hamiltonian  given by 
%--------------  (1) ----------------------
\begin{equation}
H = \sum_{i,j=1}^{N} \sum_{\alpha,\beta} t_{i,j;\alpha, \beta} |i, \alpha> <j, \beta| \; ,
\label{eq:H_model}
\end{equation}
%------------------
where  $i$ and $j$ are the  sites  of the unit cell 
  with  $N$ being the total number of the square lattice, and 
     $\alpha$ and $\beta$ denote the 8 molecular orbitals 
      given by HOMO $(H1, H2, H3, H4)$ and 
         LUMO $(L1, L2, L3, L4)$.  
 The lattice constant is taken as unity. 
There are three kinds of 
 transfer energies, $t_{i,j;\alpha, \beta}$, given by   
   HOMO-HOMO (H), LUMO-LUMO (L),  and HOMO-LUMO (HL). 

  The transfer energies 
 $t_{i,j;\alpha, \beta}$($P$)  at a pressure $P$ (GPa) 
 are estimated using 
  linear interpolation between two pressures of $P$ = 8  and 0 GPa,  
 which are given by 
%---------------  (2)  ----------------------
\begin{equation}
t_{i,j; \alpha,\beta}(P)
  = r t_{i,j; \alpha,\beta}(0) + (1-r) t_{i,j; \alpha,\beta}(P_0) \; ,
\label{eq:interpolation}
\end{equation}
 with $r = P/P_0$ and $P_0= 8$ GPa.
 The transfer energy in the unit of eV is given in Appendix.
 The pressure, where the behavior of the Dirac electron 
 has been observed experimentally, corresponds to 
 $P$=8 GPa in the first-principles calculation.\cite{Kato_JACS}

Using the  Fourier transform 
$ |\alpha(\bm{k})>= \sum_{j} \exp[- i \bm{k}\bm{r}_j] \; |j,\alpha>$
       with a wave vector  $\bk = (k_x, k_y, k_z)$, 
Eq.~(\ref{eq:H_model}) is rewritten as
%-------------- (3) ----------------------
\begin{equation}
H = 
 \sum_{\bm{k}} |\Phi(\bm{k})> \hat{H}(\bm{k}) <\Phi(\bm{k})|\; , 
\label{eq:H} 
\end{equation}
 where
the base is given by 
$<\Phi(\bm{k})| = (<H1|,<H2|,<H3|,<H4|, <L1|, <L2|, <L3|, <L4|)
 \; ,$  
 and  $\hat{H}(\bm{k})$ is  the Hermite  matrix Hamiltonian. 
 The matrix element 
$  [ \hat{H}(\bk) ]_{\alpha,\beta}$ 
is  given in Appendix.\cite{Kato_JACS,Kato2017_JPSJ}
Since the symmetry of the HOMO (LUMO) 
 is odd (even)  with respect to the 
 Pd atom, the matrix element of H-L (H-H and L-L) 
 is the odd (even) function 
 with respect to $\bm{k}$, i.e., 
an inversion symmetry  at the time reversal invariant momentum. 
The energy band  $E_j(\bk)$ 
 and the wave function $\Psi_j(\bk)$, $(j = 1, 2, \cdots, 8)$ 
 are calculated from 
%-------------- (4) ----------------------
\begin{equation}
\hat{H}(\bm{k}) \Psi_j(\bk) 
 = E_j(\bk) \Psi_j(\bk) \; , 
\label{eq:energy_band}
\end{equation}
 where $E_1 > E_2 > \cdots > E_8$ and 
%-------------- (5) ----------------------
\begin{equation}
\Psi_j(\bm{k}) = \sum_{\alpha}
 d_{j,\alpha}(\bk) |\alpha> \; ,
\label{eq:wave_function}
\end{equation}
 with $\alpha =$  
 H1, H2, H3, H4, L1, L2, L3, and L4. 
Noting that the band is  half-filled owing to the HOMO and LUMO 
bands,
 we calculate a gap function defined by 
%-------------- (6) ----------------------
\begin{equation}
E_g(\bk) = {\rm min} [E_4(\bk)-E_5(\bk)] \; ,
\label{eq:Eg}
\end{equation}
for all $\bk$ values in the Brillouin zone. 
The Dirac point $\bk_D$ is obtained from $E_g (\bkD)= 0$.

The electric conductivity is given  as follows. 
Using the component of the wave function $d_{\alpha \gamma}$ 
 in Eq.~(\ref{eq:wave_function}), 
 the response function  per spin 
%for the fixed $k_z$ 
  is calculated as\cite{Katayama2006_cond}  
%------------  (7) (8)  ----------------------
\begin{eqnarray}
\sigma^{\nu \nu'}(T) &=&  
  \frac{e^2 }{\pi \hbar N} 
  \sum_{\bk} \sum_{\gamma, \gamma'} 
  v^\nu_{\gamma \gamma'}(\bk)^* 
  v^{\nu'}_{\gamma' \gamma}(\bk) 
 \int_{- \infty}^{\infty} d \ep
   \left( - \frac{\partial f(\ep) }{\partial \ep} \right)
       \nonumber \\
 & &
 \times \frac{\Gamma}{(\ep - \xi_{\bk \gamma'})^2 + \Gamma^2} \times 
 \frac{\Gamma}{(\ep - \xi_{\bk \gamma})^2 +  \Gamma^2}
  \; ,  
  \label{eq:sigma}
\\
  v^{\nu}_{\gamma \gamma'}(\bk)& = & \sum_{\alpha \beta}
 d_{\alpha \gamma}(\bk)^* 
   \frac{\partial \tilde{H}_{\alpha \beta}}{\partial k_{\nu}}
 d_{\beta \gamma'}(\bk) \; ,
  \label{eq:v}
\end{eqnarray}
%----------------------------------
 where $\nu = x, y,$ and $z$.
 $h = 2 \pi \hbar$ and $e$ denote  Planck's constant and the electric charge, 
 respectively. 
 $\xi_{\bk \gamma} = E_{\gamma}(\bk) - \mu$ 
 and $\mu$ denotes the chemical potential obtained below.
 $f(\ep)= 1/(\exp[\ep/T]+1)$ with $T$ being temperature in the unit of eV 
 and $k_{\rm B }=1$.
 The damping constant $\Gamma$ is introduced  
 to obtain a finite conductivity  where $1/\Gamma$
  corresponds to the lifetime by  impurity scattering.
The total number of lattice sites  is given by $N=N_xN_yN_z$, 
 and the lattice constant is taken as unity.
 At low temperatures, Eq.~(\ref{eq:sigma})  is expanded as 
%-------------  (9)  (10)   -------
\begin{eqnarray}
\sigma^{\nu \nu'}(T)& =&  \sigma^{\nu \nu'}(0) 
 + C_{\nu \nu'}(T/\Gamma)^2 + \cdots \; ,
  \label{eq:sigma_T}    \\ 
C_{\nu \nu'} & = &   \frac{e^2 \pi }{12 \hbar N} 
  \sum_{\bk} \sum_{\gamma, \gamma'} 
  v^\nu_{\gamma \gamma'}(\bk)^* 
  v^{\nu'}_{\gamma' \gamma}(\bk) \Gamma^4 
     \nonumber \\
 & & \left[
   \frac{6 \xi_{\bk \gamma}^2 -2 \Gamma^2}
    {( \xi_{\bk \gamma}^2 + \Gamma^2)^3
     ( \xi_{\bk \gamma'}^2 + \Gamma^2) } \right.  
    \nonumber \\
& & + \left.
 \frac{ 4 \xi_{\bk \gamma}  \xi_{\bk \gamma'}}
    {( \xi_{\bk \gamma}^2 + \Gamma^2)^2)
     ( \xi_{\bk \gamma'}^2 + \Gamma^2)^2} \right.
      \nonumber \\
 & & + \left. \frac{6 \xi_{\bk \gamma'}^2 -2 \Gamma^2}
   {( \xi_{\bk \gamma'}^2 + \Gamma^2)^3
     ( \xi_{\bk \gamma}^2 + \Gamma^2) }
      \right]  \; ,  
  \label{eq:C_nu}    
\end{eqnarray} 
 where 
   the first term of $\sigma^{\nu \nu'}(T)$ is enough for 
   $T << \Gamma$.  
  The resistivity is given by $1/\sigma^{\nu\nu}$
 since   $\sigma^{\nu \nu'}(T)$ for 
  $\nu \not= \nu'$ is much smaller than that for $\nu=\nu'$ 
   in the present numerical calculation. 
 We mainly examine  
    the electric  conductivity at $T$=0 with $\nu=\nu'$,
   ($\nu =x, y, z$) while  $C_{\nu \nu} = C_{\nu}$ will be examined later 
 to elucidate the $T$ dependence.   
The first term of Eq.~(\ref{eq:sigma_T}) is rewritten as  
%---------------- (11) ------------------
\begin{eqnarray}
 \sigma^{\nu \nu}(0) = \sigma^{\nu}_{\rm 3D}
  &=&
     \frac{1}{N_z} \sum_{k_z} \sigma^\nu_{\rm 2D}(k_z) \; ,
  \label{eq:sigma_3D}
\end{eqnarray} 
%--------------- (12)  -----------
\begin{eqnarray}
\sigma^{\nu}_{\rm 2D}(k_z) &=&  
  \frac{e^2 }{\pi \hbar N_x N_y}
 \sum_{k_x k_y} \sum_{\gamma \gamma'}
 |v_{\gamma \gamma'}(\bk)|^2  \nonumber \\
& & \times \frac{\Gamma}{\xi_{\bk \gamma'}^2 + \Gamma^2} \times 
 \frac{\Gamma}{\xi_{\bk \gamma}^2 +  \Gamma^2}
  \; ,  
  \label{eq:sigma_2D}
\end{eqnarray} 
 where  $\sigma^{\nu}_{\rm 2D}(k_z)$ denotes the conductivity 
 on the plane with a fixed $k_z$.
A semimetal is obtained owing to the variation of the energy of the Dirac point
  $\eD(k_z)$  with respect to $k_z$.  
Equation (\ref{eq:sigma_2D}) gives  
 two-dimensional conductivity,   
 which reduces to  a universal conductivity 
 in the case of the isotropic Dirac cone  
 with  the chemical potential 
 being equal to  $\eD(k_z)$. \cite{Katayama2006_cond}
 The conductivity increases 
 when  the chemical potential
  moves  away from   $\eD(k_z)$. \cite{Suzumura_JPSJ_2014}

The chemical potential $\mu$ is determined self-consistently 
 from the half-filled condition, which is given by 
%-----------  (13)  ---------------
\begin{eqnarray}
  \frac{1}{N} \sum_{\bk} \sum_{\gamma}  f[E_{\gamma}(\bk)] = 
 \int_{-\infty}^{\infty} {\rm d} \omega D(\omega) f(\omega) =  4 \; ,  
  \label{eq:eq11}
\end{eqnarray}
where 
%----------------  (14) ------------------
\begin{eqnarray}
D(\omega) &=& \frac{1}{N} \sum_{\bk} \sum_{\gamma}
 \delta [\omega - E_{\gamma}(\bk)] \; .
  \label{eq:dos}
\end{eqnarray}
$D(\omega)$ denotes the density of states (DOS) per spin and per unit cell, 
 which satisfies  $\int {\rm d} \omega D(\omega) = 8$.
 Note that Eq.~(\ref{eq:sigma}) can be partly understood 
 using  DOS when $\gamma=\gamma'$ and the $\bk$ dependence of 
 $v_{\gamma'\gamma}^{\nu}$ close to the Dirac point is small. 
 The conductivity at $T$=0 is analyzed using $D(\omega)$.  
%----------------------------------
%%%%%%%%%%%%%%%%%%%%%%%%
\section{Conductivity of Dirac semimetal in [P$_d$(dddt)$_2$]}
%%%%%%%%%%%%%%%%%%%%%%%%

%----------------------------------
\subsection{Loop of Dirac point}
%----------------------------------
 The pressure dependence of the electronic states of [Pd(dddt)$_2$] 
is  as follows. 
At ambient pressure,   the insulating state is found where 
   there exists  a minimum  gap at $\bk=(0,0,0)$, 
    and  the LUMO bands ($E_1,\cdots, E_4$) are separated  from 
 the HOMO bands ($E_5,\cdots, E_8$).
  With increasing  pressure ($P$),   the gap at $\bk=0$  is reduced. 
 When the minimum of the LUMO band becomes smaller than 
  the maximum of the HOMO band,    
  a Dirac point emerges  between  $E_4(\bm{k})$ (conduction band)  
    and  $E_5(\bm{k})$ (valence band) at around $\bk = 0$. 
At the  Dirac point  given by   
   $E_4(\bk)= E_5(\bk)$,  
 the H-L interaction (the coupling between HOMO  and LUMO orbitals) 
  vanishes owing to  
  the orthogonality of the HOMO and LUMO wave functions.\cite{Kato_JACS}
Thus, a loop of the Dirac point is obtained at the intersection 
 of the plane of   $E_4(\bk)= E_5(\bk)$ and that of  
 the vanishing of the H-L interaction.
\cite{Kato2017_JPSJ}

%========  Fig 1 ================
\begin{figure}
  \centering
\includegraphics[width=8cm]{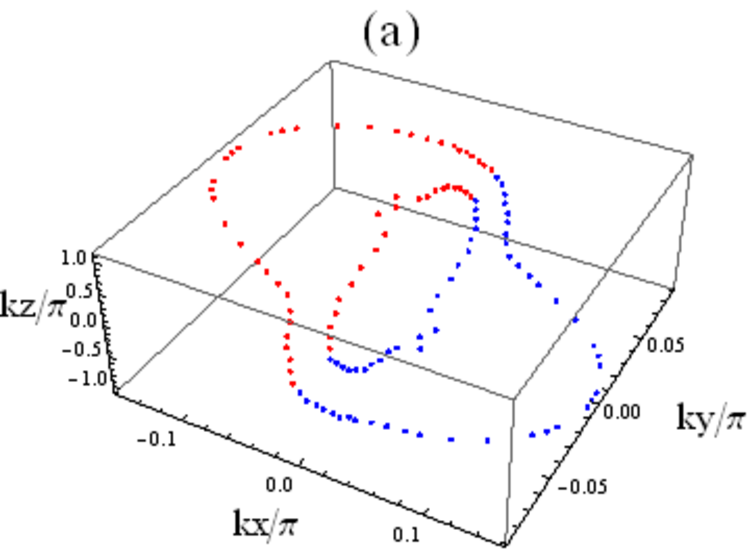}
\includegraphics[width=6cm]{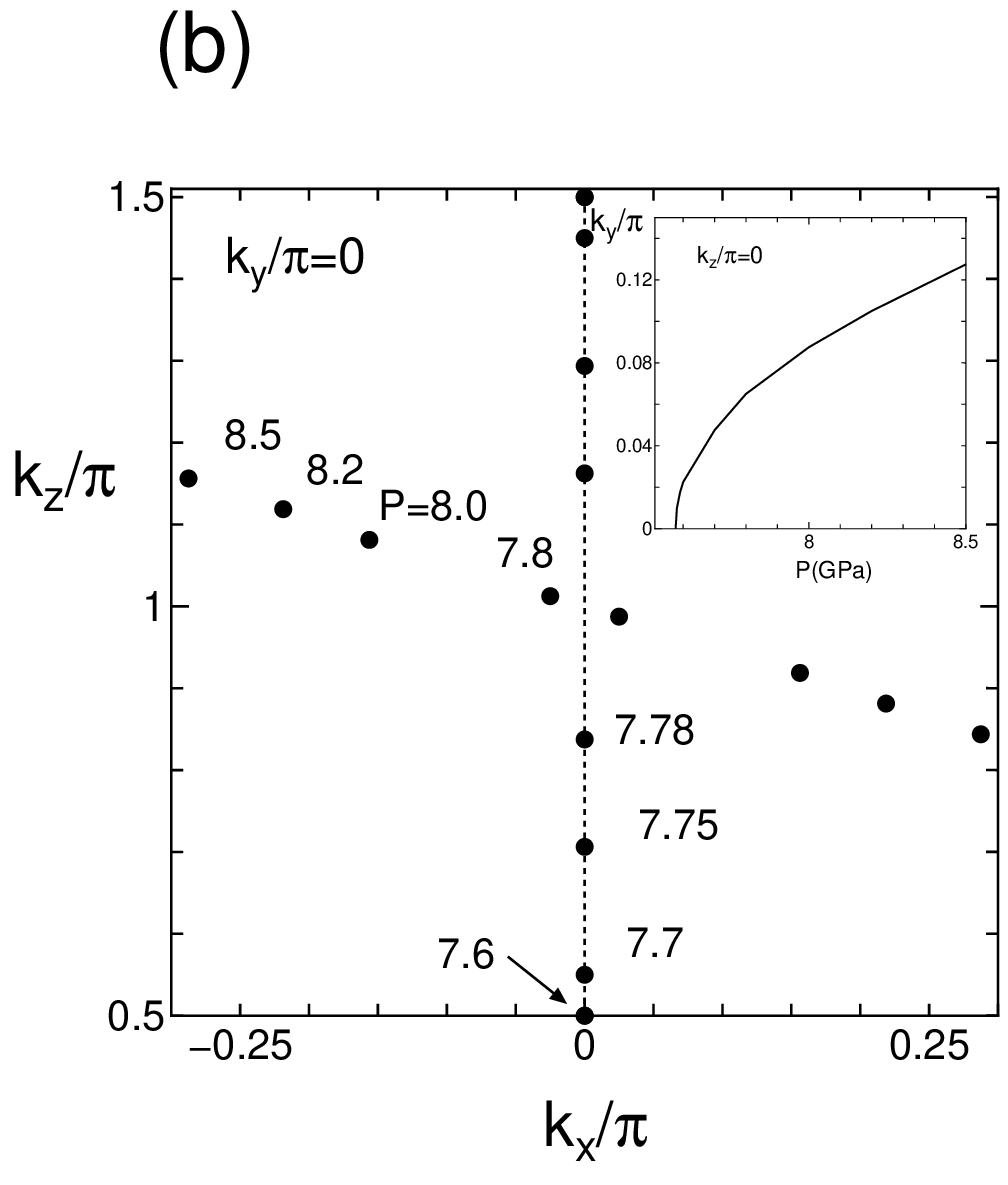}\\
\includegraphics[width=8cm]{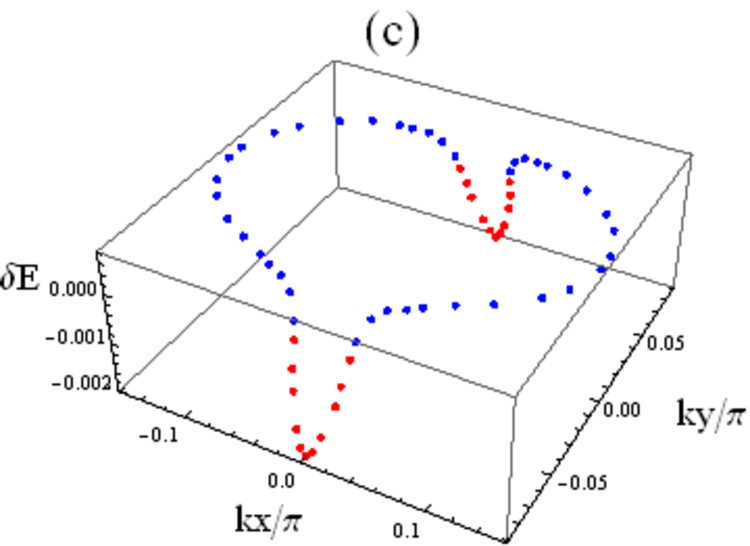}
     \caption{(Color online)
(a) 3D loop of Dirac point at P=7.8 (inside) and  8   GPa (outside). 
The Dirac point is symmetric with respect to the origin 
 $(k_x, k_y, k_z) = (0, 0, 0)$.
(b) Dirac point on the plane of $k_x$ and $k_z$ 
 where $k_x = 0$ for $7.58 < P < 7.79$
 and $|k_x| \not=0$ for $7.8 < P$.
The inset denotes the emergence of 
 the Dirac point $\bkD$   as a function of $P$ 
  where $k_x = k_z =0$ and  $k_y \not= 0$ 
 for  $P > 7.58$.  
(c) Energy  $\eD$ of the Dirac point (at  $P$=8 GPa)
   on the plane of $k_x$ and $k_y$,  
 where  $\delta E = \eD - \mu$  and   $\mu \simeq 0.5560$.  
}
\label{fig:fig1}
\end{figure}
%-------------------------------

%------------------   Fig. 1 --------------------
  Figure \ref{fig:fig1}(a) shows  the Dirac point 
 in the three-dimensional momentum space 
$\bk = (k_x, k_y,k_z)$ for P= 7.8 and 8.0 GPa,  which 
 forms  a loop in   the region of almost 
  the first Brillouin zone (inner loop) 
 and that of  the extended  zone (outer loop), respectively. 
Since the critical pressure for the loop  touching the boundary of the first 
 Brillouin zone is given by $P_c \simeq 7.79$, 
 the loop is already an extended one 
 at   the pressure of $P$ =8 GPa where the behavior of the 
    Dirac electron is found in the experiment on the resistivity.
\cite{Kato_JACS}  
In Fig.~\ref{fig:fig1}(a),
 the Dirac point for the $P$=8 GPa is   given 
 by   $\bkD/\pi=(k_x/\pi, k_y/\pi, k_z/\pi)$ =    $(0, \pm 0.0875 ,0)$ 
  and  $(\pm 1.09, 0, \pm 1)$   on the plane of 
$ k_x$-$k_y$ and $k_z$-$k_x$, respectively. 
Note that the Dirac cone is anisotropic, e.g., the ratio of 
the velocity  at $P$ =8 GPa  is estimated as 
 $v_x : v_y: v_z \simeq 1 : 5: 0.2$. 
The loop crosses both the plane of $k_z = 0$
 and $k_y = 0$. 
Figure \ref{fig:fig1}(b) shows the Dirac point on the $k_z$-$k_x$ plane 
 with some choices of pressure. 
 The Dirac point shows $k_x=0$, i.e.,  on the  $k_z$ axis   
      for $7.58 <P < 7.78$, but deviates from $k_x=0$ 
    for $7.78 < P$ with a line, 
       $k_z/\pi-1 \simeq - 1.2  k_x/\pi$. 
 The Dirac point also  exists on the line of the $k_x$ axis, 
  which is a common feature of  the loop. 
  In the  inset,   the $P$ dependence of the Dirac point 
 $\bkD [= (0, k_y, 0) ]$
 is shown where  
  $ k_y \simeq 0.0182 (P-7.58)^{1/2}$, 
   suggesting   the  emergence of the Dirac point 
  at $P \simeq$ 7.58 GPa. 

%----------------------------------
\subsection{Conductivity}
%----------------------------------
We take $e = \hbar$ = 1 in the following numerical calculation. 
 The present system exhibits  a nodal line semimetal 
  owing to the variation of the energy in the Dirac point 
    along the loop. 
 Note  that $\eD$ depends slightly on $k_z$ 
  where $\eD [ = E_4(\bk_{\rm D}) =E_5(\bk_{\rm D})]$, and 
   $E_4(\bk) > \eD (k_z) > E_5(\bk)$  for a fixed $k_z$.
 Such $\eD$ of the Dirac point at $P$ = 8 GPa is shown 
in Fig.~\ref{fig:fig1}(c) 
  corresponding to the nodal line of 
 the outside loop of Fig.~\ref{fig:fig1}(a), 
 where  $\delta E (= \eD  - \mu$), 
 and  $k_z$ is given  as the function of  $k_x$ and $k_y$. 
 For $P$=8 GPa, 
 the energy $\eD(k_z/\pi)$  as a function of  $k_z$  
 shows the relation   $\eD(0) < \mu < \eD(1.09)$, 
 where 
$\eD(0)= 0.5540 $,  
$\eD(1.09) = 0.5570 $, 
 and $\mu = 0.5560  \simeq \eD (0.65)$.
 The dispersion of $\eD(k_z)$ with a width of $\simeq$ 0.003 
  suggests 
  the electron and hole pockets  along the line of the Dirac point, 
i.e.,  a semimetallic state. 
% as shown  by DOS. 
In this section, the property of such a Dirac electron 
 is examined by calculating the electric conductivity mainly at $T$=0.

\begin{figure}
  \centering
\includegraphics[width=7cm]{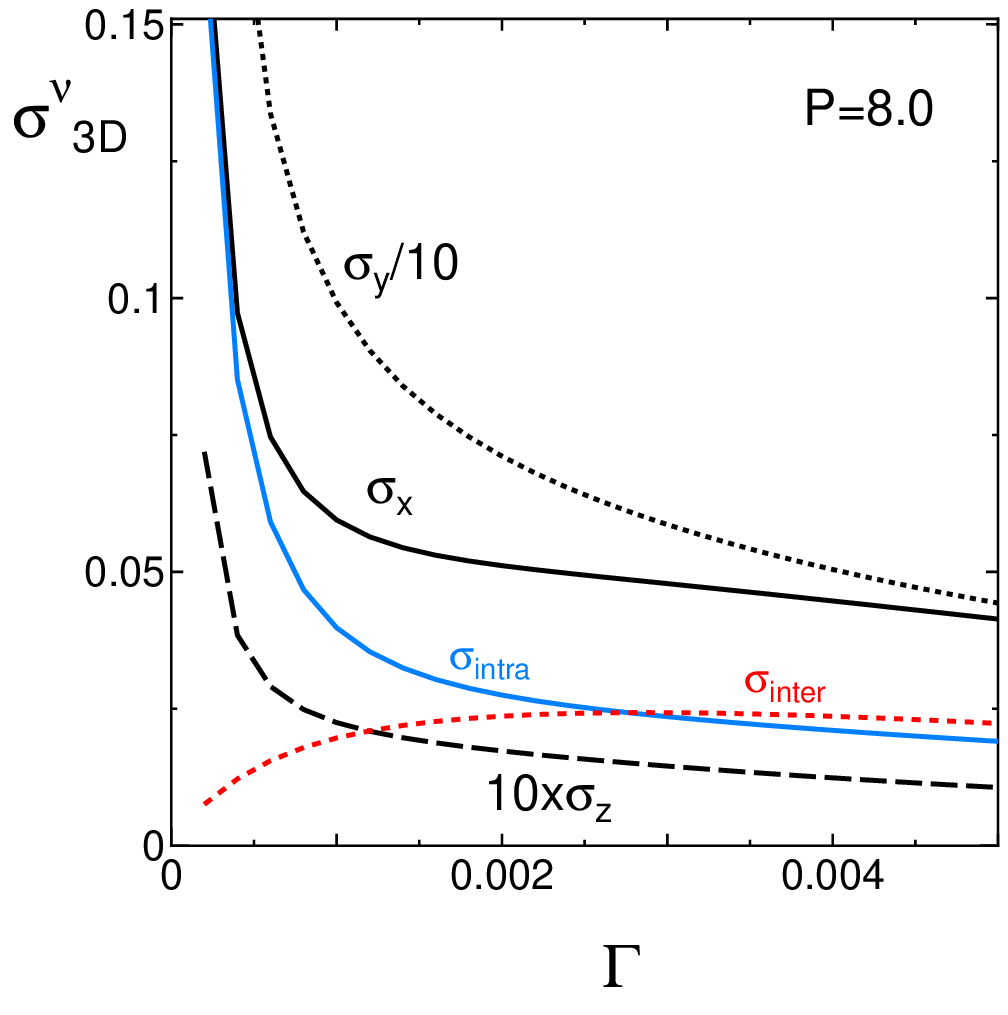} 
\caption{(Color online)
 $\Gamma$ dependence of conductivity $\sigma_{\nu}=
 \sigma^\nu_{\rm 3D}$ 
 ($\nu = x, y$, and $z$) at  $P$ = 8 GPa, 
 where 
   $\sigma_y > \sigma_x > \sigma_z$. 
The interband and intraband components are also shown  
 for $\sigma_x (= \sigma_{\rm intra} + \sigma_{\rm inter})$.
}
\label{fig:fig2}
\end{figure}

%---------   Fig. 2  ---------
In Fig.~\ref{fig:fig2},  
 $\sigma_{\nu} (=\sigma^\nu_{\rm 3D})$ 
  for  $P$= 8 GPa  is shown  as a function of $\Gamma$.
The rapid  increase in $\sigma_\nu$ 
  with decreasing $\Gamma$  occurs  
 as a competition between $\Gamma$ and 
 the deviation of $\mu$ from $\eD$.
In fact, the increase in $\sigma_\nu$ occurs  
 when $|\mu - \eD|$   becomes larger than $\Gamma$ for a single Dirac cone.
\cite{Suzumura_JPSJ_2014}
 In the case of Fig.~\ref{fig:fig2}, the increase is found for $\Gamma$ being 
 smaller than $<|\delta \mu|>_{\rm av}$, 
 where $<|\delta \mu|>_{\rm av}$  denotes the average of 
 $|\eD(k_z)-\mu|$ with respect to $k_z$. 
The characteristic of the Dirac electron where  $\sigma_\nu$ becomes constant 
 is seen for a large $\Gamma$.
\cite{Suzumura_JPSJ_2014}
 Another aspect of the Dirac electron is seen  from the component of 
 $\sigma_\nu$, which consists of 
  the interband contribution  ($\gamma \not= \gamma'$)
 and  intraband  contribution  ($\gamma=\gamma'$).
Note that $\gamma, \gamma'$ = 4 and 5 is sufficient in the present calculation. 
When $<|\delta \mu|>_{\rm av}$ is  larger than $\Gamma$, 
 the intraband contribution is dominant.
When $<|\delta \mu>_{\rm av}|$ is much smaller  than $\Gamma$, 
 the interband contribution becomes nearly equal to  
  the intraband contribution. 
%due  to   $\xi_{\gamma}(\bk) \simeq - \xi_{\gamma'}(\bk)$ in  Eq.~(\ref{eq:sigma}).
Such a crossover from small $\Gamma$ to large $\Gamma$ 
 is shown for $\sigma_x$, where  
 $\sigma_{\rm intra}$ is compared with 
 $\sigma_{\rm inter}$.  
Thus, in the present calculation of the conductivity, 
   we take $\Gamma$  = 0.002, which is a reasonable  magnitude to find 
  the Dirac behavior in  the 
 tight-binding model of [Pd(dddt)$_2$].

%---------------  Fig 2 --------------------
 %------------------

%---------------  Fig. 3   --------------------
\begin{figure}
  \centering
\includegraphics[width=7cm]{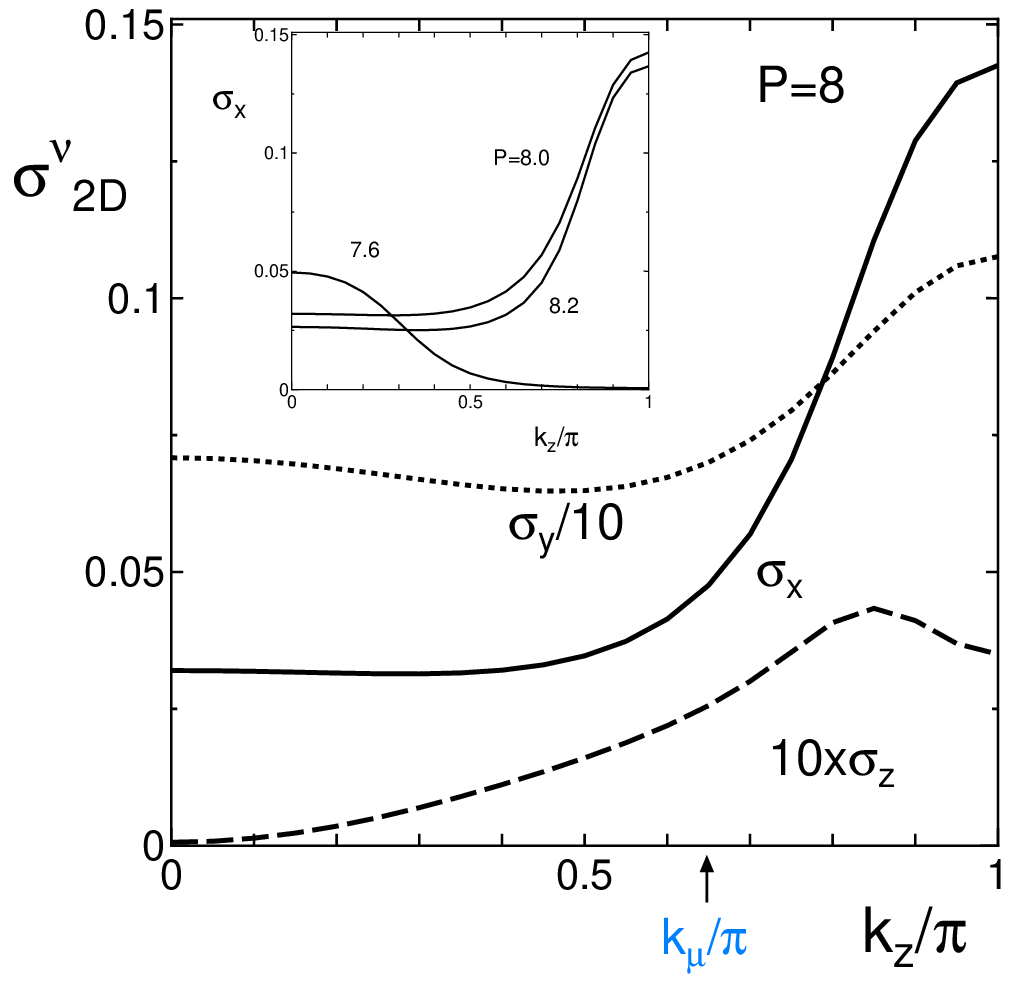}   
\caption{(Color online)
 $k_z$ dependence of conductivity $\sigma_{\nu}=
 \sigma^\nu_{\rm 2D}$ 
 ($\nu = x, y$, and $z$) at  $P$ = 8 GPa, 
 where 
   $\sigma_y > \sigma_x > \sigma_z$. 
 $\Gamma = 0.002$.
$k_\mu$ denotes $k_z$  corresponding to $\eD(k_\mu) =\mu$. 
The inset shows 
 a comparison of $\sigma_x(k_z)$ for P=8 
 with those  for P = 7.6 and 8.2, which 
 correspond to the loop  within the first Brillouin zone  
 and extended Brillouin zone. 
}
\label{fig:fig3}
\end{figure}

The conductivity of the nodal line semimetal is essentially  determined by 
 electrons close to  the  loop of the Dirac point.
To comprehend the contribution of the respective location of 
the loop,  we examine  $\sigma^\nu_{\rm 2D}(k_z)$ 
 given by Eq.~(\ref{eq:sigma_2D}),
which 
shows two-dimensional behavior  associated with  the Dirac cone 
 on the  plane of fixed $k_z$. 
Figure \ref{fig:fig3} shows $\sigma_\nu =\sigma^\nu_{\rm 2D}(k_z)$ 
  for $P$= 8 GPa  as a function of $k_z$ 
 in the reduced  zone, i.e., the loop is folded at $k_z=\pi$. 
The  constant behavior of $\sigma_x$ for $k_z/\pi <0.5$  suggests 
 that the Dirac cone is robust on the $k_x$-$k_y$ plane,  
 and the increase in $\sigma_x$  for $0.6 < k_z/\pi <1$ occurs owing to   the rotation of the cone into the $k_z$-$k_x$ plane [outer loop shown by Fig.~\ref{fig:fig1}(a)], resulting in an increase in  the density of the Dirac point per $k_z$. 
 Note that the ratio of $\sigma^\nu_{\rm 2D}(\pi)/\sigma^\nu_{\rm 2D}(0)$ 
  for $\nu = x$ 
 is larger than that for $\nu =y$.
From  an effective model of the 2 x 2 Hamiltonian of the Dirac cone with 
zero doping,\cite{Suzumura_JPSJ_2014}
  the ratio of the conductivity is given by 
 $\sigma_x/\sigma_y \simeq (v_x/v_y)^2$,  where $v_x$ and $v_y$ 
  are  the  velocities  of the Dirac cone 
     along $k_x$ and $k_y$ directions, respectively.
The Dirac cone  on the $k_x$-$k_y$ plane at  $k_z/\pi = 0$ shows 
 the velocity $v_x \simeq 0.016$ eV and $v_y \simeq 0.10$ eV, 
 while the cone  on the $k_z$-$k_x$ plane 
  at  $k_y/\pi = 0$ 
 shows  $v_x \simeq 0.019$ eV and $v_z \simeq 0.0037$ eV. 
 These velocities are estimated from the gap function of $E_g(\bk)$ 
 shown in Figs.~2(b) and 2(d) of Ref.~\citen{Suzumura2017_JJAP}. 
For $k_z \simeq 0$,  the ratio  $\sigma_y/\sigma_x$ 
 obtained from Fig.~\ref{fig:fig3} 
 (dotted and solid lines) is about  half of  that expected from the effective model. This difference may come from the fact that
  the chemical potential is located above $\eD$, i.e., 
  corresponding to the apex of the cone. 
For $k_z/\pi \simeq  0.91$ (i.e., 1.09 corresponding to the extended zone)
 the ratio of $\sigma_x/\sigma_z$   obtained from Fig.~\ref{fig:fig3} 
 (solid and dashed lines) is nearly  equal to that expected from the effective
 model of  the Dirac cone  on the $k_z$-$k_x$ plane. 
Thus, it turns out  that $\sigma_x$  exhibits a reasonable behavior of 
  the nodal semimetal since $\sigma_x$ comes from 
  all the Dirac points on the loop.  
The inset denotes $\sigma_x(k_z)$ for $P$ = 7.6, 8.0, and 8.2. 
The peak of $\sigma_x$ for $P$=7.6 is seen at $k_z=0$ 
since the corresponding loop is small and located at around $\bk = 0$.
For $P$=8 and 8.2, the peak is seen at $k_z = \pi$ owing to the loop in the extended Brillouin zone. 
The fact that   $\sigma_x$ at $P$=8.2 is smaller than that at $P$=8      
 for an arbitrary $k_z/\pi$ suggests a reduction in  the density of states, 
  as will be shown later.

%---------------  Fig.4   --------------------
\begin{figure}
  \centering
\includegraphics[width=7cm]{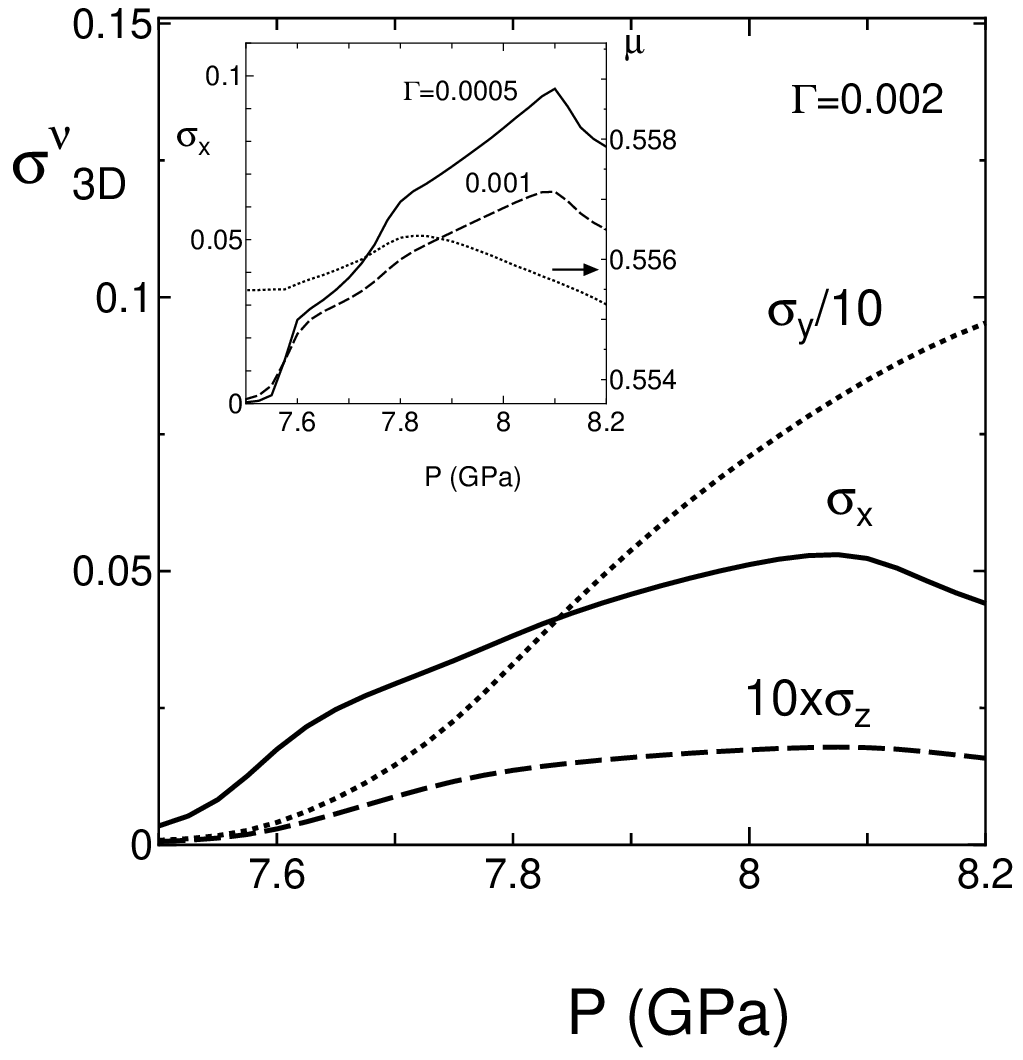}   
\caption{%(Color online)
Pressure dependences of 
 $\sigma^\nu_{\rm 3D} (= \sigma_\nu)$ for $\sigma_x$, $\sigma_y$ and $\sigma_z$ for $\Gamma$ =0.002.
$\sigma_y > \sigma_x > \sigma_z$.
The inset shows the $P$ dependences of 
 $\sigma_x$ for $\Gamma$ = 0.0005 (solid line) and 0.001 (dashed line),  
 and that of the chemical potential  $\mu$ (dotted line). 
 }
\label{fig:fig4}
\end{figure}

%===========================================================

In Fig.~\ref{fig:fig4},
 the pressure dependence of  $\sigma_\nu (=\sigma^\nu_{\rm 3D} )$ 
 ($\nu = x, y,$ and $z$) is shown  with $\Gamma = 0.002$. 
The  relation  $\sigma_y > \sigma_x > \sigma_z$ is the same as 
 that in  Fig.~\ref{fig:fig3}. 
At $P$=7.5, 
  $\sigma_y$ and $\sigma_z$ reduce sufficiently 
 but a small $\sigma_x$ remains where  
  $\Gamma$  is  smaller than the insulating gap  $E_g \simeq 0.004$ for 
 $P$  =7.5. 
With increasing $P (> 7.58)$, $\sigma_\nu$ increases owing to the increase of the loop.
 The increase of $\sigma_y$ is much larger than those of 
  $\sigma_x$ and $\sigma_z$.
A crossover from the insulating state to the semimetallic state is seen 
 in  $\sigma_x$ for  $P \sim 7.6$, 
 while $\sigma_y$ and $\sigma_z$ show a monotonic increase. 
  Note that   $\sigma_x$ takes a maximum at $P \sim$ 8.1.
 The inset denotes the $P$ dependences of  $\sigma_x$ 
 for $\Gamma$ = 0.0005 and 0.01 and that of the chemical potential $\mu$. 
  For a smaller $\Gamma$, the magnitude of the peak 
  becomes larger,  as is also seen from  Fig.~\ref{fig:fig2}. 
 Such  characteristic pressure ($\simeq$ 8.1) for the peak 
is almost independent of  $\Gamma$, and the peak moves to a cusp in the limit of  small $\Gamma$. 
Since $<|\delta \mu|>_{\rm av}$ is  larger than $\Gamma$, 
 such a peak is attributable to the intraband 
contribution suggesting  a significant role of the  nodal line semimetal. 
The chemical potential also shows a maximum at $P \sim 7.8$   and decreases 
linearly 
 with  further increase in pressure, e.g., $\mu \simeq 0.5540$  at $P$ = 8.5.

%---------------  Fig. 5   --------------------
\begin{figure}
  \centering
\includegraphics[width=7cm]{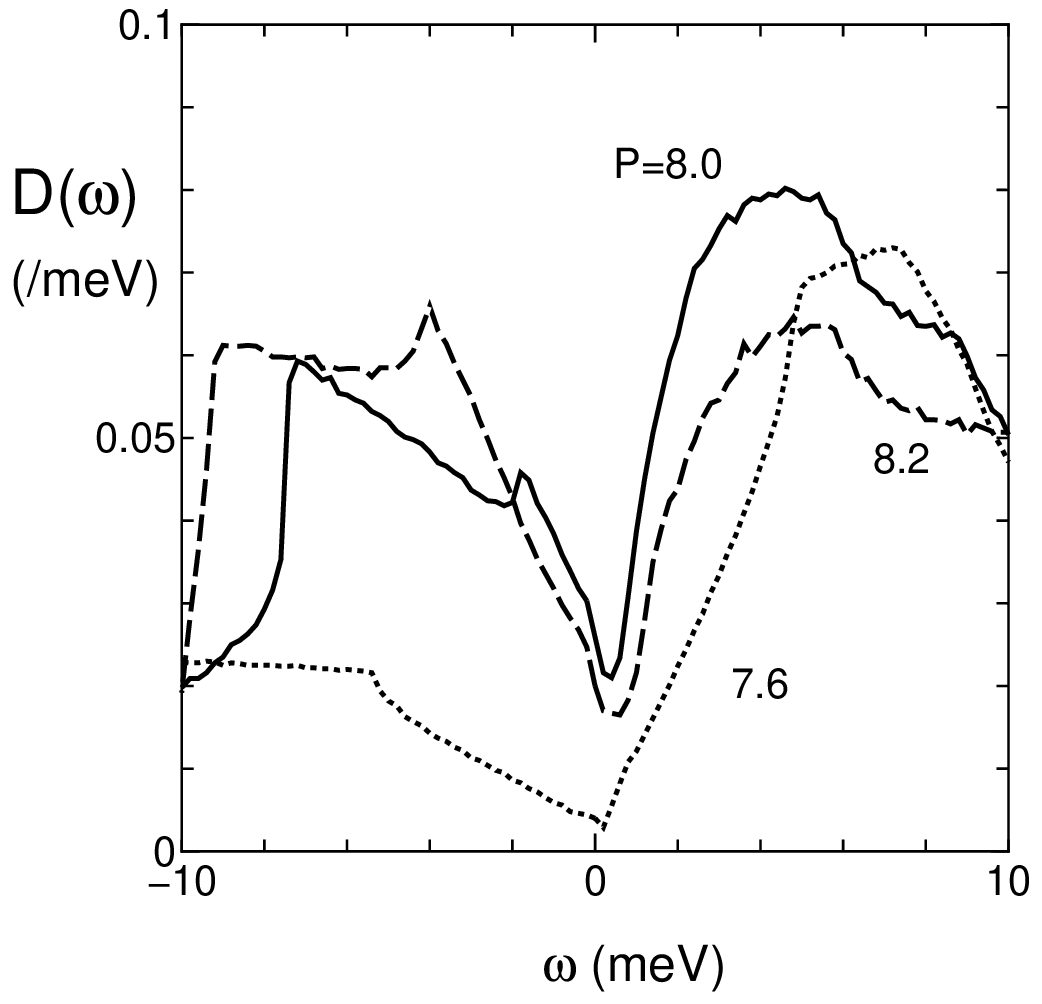}   
\caption{%(Color online)
Density of states $D(\omega)$ for 
$P$= 7.6, 8, and 8.2 
 where the origin is taken as the respective chemical potential. 
}
\label{fig:fig5}
\end{figure}
%----------------------------------------------
 To examine the maximum $\sigma^x_{\rm 3D}$, 
 we calculate the density of states $D(\omega)$ given by Eq.~(\ref{eq:dos}). 
 Figure \ref{fig:fig5} shows  $D(\omega)$ for the fixed 
 $P$=7.8, 8.0, and 8.2, where $\omega$ = 0 corresponds to the respective 
 chemical potential. 
 The minimum  $D(\omega)$ is located slightly above the chemical potential. 
 At lower pressures, insulating behavior is obtained, e.g., 
$D(0) = 0$ for $P < 7.58$ and a gap is given 
 by $E_g \simeq 0.004$  at $P$=7.5. 
At  $P$ = 7.6,  $D(\omega)$  with  $\omega < 0$ 
 is smaller than that with  $\omega > 0$. 
 Such a difference reduces with increasing pressure.
At $P$ = 8.2, $D(\omega)$ with  $\omega > 0$  decreases 
 in addition to the decrease in $D(0)$, suggesting 
 optimum  electrons close to the chemical potential 
 at around $P$ = 8.  
 In fact,
  the quantity given by   $\int D(\omega) d \omega$ in the interval range of
  $-0.002 < \omega <0.002$ 
 is large   for  $P$=8.0 compared with that for $P$=7.6 and 8.2.
A  quantity that is estimated 
 from   Eqs.~(\ref{eq:sigma_3D}) and (\ref{eq:sigma_2D}) without $v_{\gamma \gamma'(\bk)}$ also shows  a maximum at $ P \simeq  7.9$.
Thus, the pressure dependence of $\sigma^x_{\rm 3D}$ is reasonably understood 
 in terms of DOS  except for  the interband contribution.  
 By noting a similar  peak  for the $P$ dependences of  $D(0)$  and  $\mu$, the  peak of $\sigma^x_{\rm 3D}$ 
  at $P \simeq 8.1$, which is slightly larger than that of $D(0)$ and $\mu$, 
 may be a  combined effect 
 of  $v_{\gamma \gamma'(\bk)}$ and   a factor given by 
$(\xi_{\bk \gamma'}^2 + \Gamma^2)^{-1}
( \xi_{\bk \gamma}^2 + \Gamma^2)^{-1}$ in Eq.~(\ref{eq:sigma})
 where the former increases slightly  with increasing $P$.
%latter factor comes is relevant to    $D(\omega)$ for $\gamma = \gamma'$.  

%---------------  Fig. 6   --------------------
\begin{figure}
  \centering
\includegraphics[width=7cm]{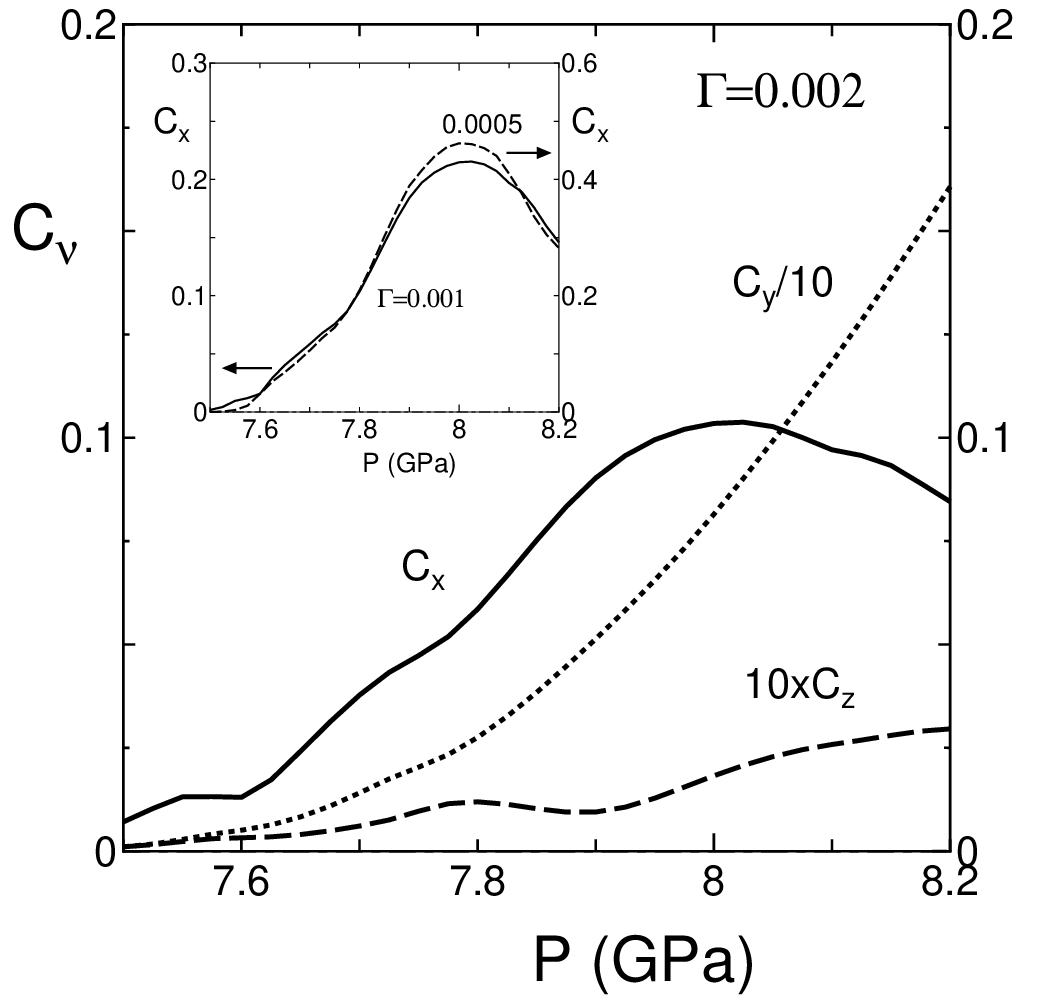}   
\caption{%(Color online)
Pressure dependence of coefficient of $C_{\nu}$ for 
 $\Gamma = 0.002$ where 
$\sigma_{\nu}(T) = \sigma_{\nu}(0) + C_{\nu}(T/\Gamma)^2$
 and $T$ denotes temperature in the unit of eV. 
The second term becomes comparable to  the first term for $T \simeq \Gamma$.
The inset denotes the corresponding $C_x$
 for $\Gamma$ = 0.001 (solid line) and 0.0005 (dashed line).
}
\label{fig:fig6}
\end{figure}
%----------------------------------------------
Finally we examine the  temperature ($T$) dependence of 
  $\sigma^\nu_{\rm 3D}$, 
 which shows an increase  proportional 
  to $T^2$,  as shown by Eq.~(\ref{eq:C_nu}).
Figure \ref{fig:fig6} shows  the pressure dependence of the 
 coefficient $C_\nu$ 
 where the relation  $C_y > C_x > C_z$ is the same  as that of $\sigma_\nu$. 
 Similarly to  $\sigma^x_{\rm 3D}$,  
   there is a maximum  
  $C_x$ at $P \simeq 8$, and  $C_y$ shows a monotonical increase. 
Comparing Fig.~\ref{fig:fig6} with Fig.~\ref{fig:fig4}, it is found that 
  the first term of Eq.~(\ref{eq:sigma_T})  becomes comparable to 
  the second term of Eq.~(\ref{eq:sigma_T}) for $T \simeq \Gamma$. 
 Thus,  $\sigma_\nu \sim {\rm const.}$ (i.e., 
  independent of $T$) is expected for  $T \ll \Gamma$.    
In the inset,   $C_x$  corresponding  to  $\Gamma$ = 0.001 (solid line) 
 and 0.0005 (dashed line) is shown for  comparison.
 There is a similarity in the $P$ dependence  of $C_x$ 
  between these three $\Gamma$'s
  since 
 the ratio of $C_x$ at $\Gamma$ = 0.002, 0.001, and 0.0005 is 
 approximately given by  2:1:0.5.  
  The  second term of  Eq.~(\ref{eq:sigma_T}) becomes  larger 
  for a smaller $\Gamma$. 
 This behavior, which is similar to 
 the $\Gamma$ dependence of   $\sigma_x$ in Fig.~\ref{fig:fig2}, 
 can be  understood qualitatively using  
  the effective model of a 2 x 2 Hamiltonian 
of the Dirac cone with  the linear dispersion.
\cite{Suzumura_JPSJ_2014}

%================   Summary ===========================
\section{Summary and Discussion}
We examined electric  conductivity  for a nodal line semimetal 
 in [Pd{dddt)$_2$] under pressure 
    using a tight-binding model. 
The semimetal comes from the chemical potential located on 
  the loop of the Dirac point 
    between the conduction and valence bands [$E_4(\bk)$ and $E_5(\bk)$].  
From 
  the gap function  $E_g(\bk) = E_4(\bk)- E_5(\bk)$, 
 where  the Dirac point is given by  $E_g(\bkD)=0$, 
 the velocity of the Dirac cone is estimated 
   to understand electric conductivity, which 
    is determined by electrons near the  loop.
 There is a large anisotropy of the conductivity given by 
 $\sigma_y > \sigma_x > \sigma_z$, 
    where  $y$ denotes the direction of the stacking of  molecules 
     and $z$ is the  interlayer direction. 
 The contour of $E_g(\bk)$  is an  ellipse with anisotropic velocities of 
 the Dirac cone, e.g., 
  $v_x \simeq 0.016$ eV and $v_y \simeq 0.10$ eV at $k_z/\pi$ = 0
 ($v_x \simeq 0.019$ eV and $v_z \simeq 0.0037$ eV at $k_y/\pi$ = 0). 
 The anisotropy of the conductivity 
  comes from  the  anisotropy of the velocity, e.g.,
  $\sigma_x/\sigma_y \sim v_x^2/v_y^2$. 
The moderate magnitude of the  damping constant $\Gamma$ is taken  
 to obtain a finite conductivity. 
The constant  resistivity at low  temperature ($T$) 
 is expected  from the coefficient $C_x$ [Eq.~(\ref{eq:C_nu})]
 when $T < \Gamma$.
 The pressure ($P$ GPa) dependence of conductivity was evaluated in the 
 interval range of $7.6 < P <8.2$ 
 to show a maximum  at  $P \simeq 8.1$.
Such  behavior of the conductivity was examined using 
  the density of states  around the chemical potential. 
 The maximum, which is associated  
 with a loop of the  Dirac point, comes from the fact that 
  the  loop increases   but 
 the density of states  decreases with increasing pressure. 

 Here, we note the energy dependence of the  nodal line located 
 around the chemical potential, which is shown in Fig.~\ref{fig:fig1}(c). 
There is a noticeable difference in the energy dependence 
 between   
 $\eD  > \mu$ and  $\eD  < \mu$.     
This  may be useful  for further studies on the orbital magnetization 
 in the present material, 
  which is  a possible topological (quasi) response inherent to 
 a nodal line.\cite{Ramamurthy2017}

Finally, we comment  on the experiment
 where  the resistivity  becomes  
  independent of temperature  under  a  given pressure. 
 This behavior may be  relevant to the Dirac electron 
  since  such a temperature independence of electric  conductivity 
 at low temperatures is  also obtained  
 theoretically  for  the  nodal  line semimetal of [Pd(dddt)$_2$] as shown 
 by Eq.~(\ref{eq:sigma_T}) and in Fig.~\ref{fig:fig6}.  

%
%-----------------------
\acknowledgements
%----------------------
The author thanks  R. Kato for useful discussions, 
 and  M. Ogata for useful comments. 
This work was supported 
 by JSPS KAKENHI Grant Number JP15H02108.

%\newpage

%==========================================
\appendix
 \section{Matrix elements of Hamiltonian  $ \hat{H}(\bm{k})$ }
On the basis of 
$<\Phi(\bm{k})| = (<H1|,<H2|,<H3|,<H4|, <L1|, <L2|, <L3|, <L4|)
 \; ,$
which is rewritten as 
$(<1|,<2|,<3|,<4|, <5|, <6|, <7|, <8|)$,
 the matrix elements $[\hat{H}(\bm{k})]_{i,j}
 = h_{i,j}$ ($i,j =1, \cdots 8$) 
 in  Eq.~(\ref{eq:H}) are 
 obtained as\cite{Kato_JACS,Kato2017_JPSJ}  
\begin{eqnarray}
h_{1,1} &=& h_{3,3} =
2 b_{1H} \cos \kb \; ,   \\ 
h_{1,2} &=& a_H(1+\e^{-i(\kabc)})\; ,   \\ 
h_{1,3}  &=&  p_H(1+\e^{-i\kb}+\e^{-i\ka}+\e^{-i(\kab)}) \; , \\ 
h_{1,4}  &=&  c_H(1+\e^{i\kc})\; , \\ 
h_{1,5}  &=&    b_{1HL}(\e^{i\kb}-\e^{-i\kb}) \; ,\\ %
%t_{1,6}  &=&  0 \; ,\\ %
h_{1,7}  &=&  p_{1HL}+ p_{2HL}\e^{-i\kb}-p_{2HL}\e^{-i\ka} 
  \nonumber \\   & & 
           -p_{1HL}\e^{-i(\kab)}  \; ,\\ %
%t_{1,8}  &=& 0 \; ,\\ %
h_{2,2}  &=& h_{4,4} =  2 b_{2H} \cos \kb  \; ,\\ %
h_{2,3}  &=&  c_H(1+\e^{i\kc}) \; ,\\ %
h_{2,4}  &=&  q_H(\e^{i(k_x+k_z)}+\e^{i(\kabc)} \nonumber \\
  & & +\e^{i\kc}+\e^{i(\kbc)})\; ,\\ %
h_{2,5}  &=&   a_{HL}(1-\e^{i(\kabc)})\; ,\\ %
h_{2,6}  &=&   b_{2HL}(\e^{ik_y}-\e^{-ik_y}) \; ,\\ %$ %\nonumber \\ 
h_{2,7}  &=&  c_{HL}(\e^{-ik_y}-\e^{i(\kbc)}) \; ,\\ %
h_{2,8}  &=&  q_{1HL}\e^{i(\kac)}+q_{2HL}\e^{i(\kabc)} 
  \nonumber \\   & & 
 - q_{2HL}\e^{i\kc}-q_{1HL}\e^{i(\kbc)}\; ,\\ %
%h_{3,3}  &=&  2 b_{1H} \cos \kb \; ,\\ %
h_{3,4}  &=& a_H(\e^{i\kb}+\e^{i(k_x+k_z)}) \; ,\\ %
h_{3,5}  &=&   p_{2HL} + p_{1HL}\e^{i\kb} \nonumber \\
 & & -p_{1HL}\e^{i\ka}-p_{2HL}\e^{i(\kab)} \; ,\\ %
%t_{3,6}  &=&  0 \; ,\\ %
h_{3,7}  &=&  b_{1HL}(\e^{i\kb}-\e^{-i\kb})\; ,\\ %
%t_{3,8}  &=&  0 \; ,\\ %
%h_{4,4}  &=&  2 b_{2H} \cos \kb \; ,\\ %
h_{4,5}  &=&   c_{HL}(\e^{-i\kb}-\e^{i(\kb-\kc)})\; ,\\ %
h_{4,6}  &=&  q_{2HL}\e^{-i(\kac)}+ q_{1HL}\e^{-i(\kabc)}
  \nonumber \\   & & 
-q_{1HL}\e^{-i\kc} -q_{2HL}\e^{-i(\kbc)}\; ,\\ %
h_{4,7}  &=&  a_{HL}(\e^{-i\kb}-\e^{-i(\kac)}) \; ,\\ %
h_{4,8}  &=&  b_{2HL}(\e^{i\kb}-\e^{-i\kb}) \; ,\\ %
h_{5,5}  &=& h_{7,7} =  \Delta E + 2 b_{1L} \cos \kb \; ,\\ %
h_{5,6}  &=&  a_{L}(1+\e^{-i(\kabc)})\; ,\\ %
h_{5,7}  &=&  p_{L}(1+\e^{-i\kb}+\e^{-i\ka}+\e^{-i(\kab)})\; ,
  \nonumber \\
   \\ 
h_{5,8}  &=&  c_{L}(\e^{i\kb}+\e^{i(-\kb + \kc)}) \; ,\\ %
h_{6,6}  &=& h_{8,8} = \Delta E + 2 b_{2L} \cos \kb \; ,\\ %
h_{6,7}  &=&   c_{L}(\e^{-i\kb}+\e^{i(\kbc)})\; ,\\ %
h_{6,8}  &=&  q_{L}(\e^{i(\kac)}+ \e^{i(\kabc)} \nonumber \\
  & & +\e^{i\kc}+\e^{i(\kbc)}) \; ,\\ %
%h_{7,7}  &=& \Delta E + 2 b_{1L} \cos \kb \; ,\\ %
h_{7,8}  &=& a_{L}(\e^{i\kb}+\e^{i(k_x+k_z)})\; ,%
%h_{8,8}  &=&  \Delta E + 2 b_{2L} \cos \kb \; , %
\label{eq:matrix_element} 
\end{eqnarray}
 where $h_{1,6}=h_{1,8}=h_{3,6}=h_{3,8}=0$, and 
$h_{i,j}=h_{j,i}^*$.
 We take  eV as the unit of energy.
The transfer energies  $t_{i,j;\alpha, \beta}$ in $h_{i,j}$  
 are calculated from an interpolation formula,
  Eq.~(\ref{eq:interpolation}),
 where  
  transfer energies at the following two  pressures 
 have been  calculated using  
 the extended  H\"uckel method. 
  At  a pressure of  $P$ = 8 GPa\cite{Kato_JACS},  
( 0 GPa \cite{Kato2017_JPSJ}) 
 $t_{i,j;\alpha, \beta}$ 
is given by 
$a_{H}=-0.0345 (-0.0136) $, 
$a_{L}=-0.0 (-0.0049) $,
$a_{HL}=0.0260 (0.0104 ) $,
$b_{1H}=0.2040 ( 0.112 ) $,
$b_{1L}=0.0648 (0.0198) $,
$b_{1HL}=0.0219 ( 0.0214) $,
$b_{2H}=0.0762 (0.0647 ) $,
$b_{2L}=-0.0413 (0.0) $,
$b_{2HL}=-0.0531 (-0.0219 ) $,
$c_{H}=0.0118$ (0.0 ),
$c_{L}=-0.0167$ (-0.0031),
$c_{HL}=0.0218$ (0.0040 ),
%-------------------------
$p_{H}=0.0398 (0.0102 ) $,
$p_{L}=0.0205 (0.0049 )$,
$p_{1HL}=-0.0275 (-0.0067 )$,
$p_{2HL}=-0.0293 (-0.0074 )$,
$q_{H}=0.0247 (0.0067 )$,
$q_{L}=0.0148 (0.0037 )$,
$q_{1HL}=-0.0186 (-0.0048 )$, and 
$q_{2HL}=-0.0191 (-0.0051 )$. 
The gap between the energy of HOMO and that of LUMO is taken 
 as $\Delta E = $ 0.696 eV to reproduce 
  the energy band by first-principles calculation.

%========================================

%========================================
\end{document}